\documentclass[12pt,preprint,]{article}
\usepackage{graphicx}
\usepackage[cp1251]{inputenc}
\usepackage{epsfig}
\usepackage[english]{babel}

\date{}
\def\b{\begin{equation}}
\def\e{\end{equation}}
\def\bee{\begin{enumerate}}
\def\eee{\end{enumerate}}
\def\be{\begin{vmatrix}}
\def\ee{\end{vmatrix}}

\begin{document}
\setcounter{page}{1}
\pagestyle{plain}

\title{\bf{ Laguerre polynomials method in valon model }}

\author{G.R.Boroun \thanks{E-mail address: boroun@razi.ac.ir, grboroun@gmail.com.}, M.Amiri}
\maketitle{\centerline{Department of Physics, Razi University,
Kermanshah 67149, Iran}

\begin{abstract}
We used the Laguerre polynomials method for determination of the
proton structure function in the valon model. We have examined the
applicability of the valon model with respect to a very elegant
method, where the structure of the proton is determined by
expanding valon distributions and valon structure functions on
Laguerre polynomials. We compared our results with the
experimental data, GJR parameterization and DL model. Having
checked, this method gives a good description for the proton
structure function in valon model.\end{abstract}

\vspace{0.5cm} {\it \emph{Keywords}}: Valon model, Structure
function, Laguerre polynomials.

\newpage
\section*{1. Introduction}
Structure functions in lepton-nucleon deep-inelastic scattering
(DIS) are the established observables probing Quantum
Chromodynamics (QCD), the theory of the strong interaction, and in
particular the structure of the nucleon. The structure functions
provide unique information about the deep structure of the hadrons
and most importantly, they form the backbone of our knowledge of
the parton densities. In QCD, structure functions are defined as
convolution of the universal parton momentum distributions inside
the proton and coefficient functions, which contain information
about the boson-parton interaction. The parton distributions in
proton have been studied extensively in a wide range of both $x$
and $Q^{2}$, as they are accurate but inconvenient to describe
analytically. Here we elaborate on the valon model that can be
very useful in the study of hadronic structure, in particular when
the experimental data are scarce. A valon has its own cloud of
partons which can be calculated in pQCD. This structure is
universal and independent of the hosting hadron [1-2].\\
The valon model [1-4] is a phenomenological model which is proven
to be very useful in its application to many areas of the hadron
physics. A valon is defined to be a dressed valence quark in QCD
with a cloud of gluons and sea quarks and antiquarks. Its
structure can be resolved at high enough $Q^{2}$ probes. In the
scattering process the virtual emission and absorption of gluon in
a valon becomes bremsstrahlung and pair creation, which can be
calculated in QCD. At sufficiently low $Q^{2}$ the internal
structure of a valon cannot be resolved and hence, it behaves as a
structureless valence quark. At such a low value of $Q^{2}$, the
nucleon is considered as bound state of three valons, UUD for
proton. The binding agent is assumed to be very soft gluons or
pions.  Let us denote the distribution of a valon in a nucleon by
$G_{\frac{v}{N}}(y)$ for each valon $v$. It satisfies the
normalization condition, $\int_{0}^{1}G_{\frac{v}{N}}(y)dy=1$ and
the momentum sum rule
$\sum_{v}\int_{0}^{1}G_{\frac{v}{N}}(y)dy=1$, where the sum runs
over all valons in nucleon $N$. The organization of the paper is
as follows; in section 2 we will give a brief description of the
valon model. In section 3 we consider the Laguerre polynomials
method which this work is based on, then we will calculate the
nucleon structure in terms of the valon distributions and the
valon structure functions on Laguerre polynomials. Finally, in
section 4,  the numerical calculations will be outlined; then we
discuss some qualitative implications of the Laguerre polynomials
method on the structure of the nucleon in valon model.\\

\section*{2. Valon distribution and Valon structure function}

In the preceding section, we discussed the valon structure of a
nucleon. In this subsection we will give the distributions in a
valon. The nucleon structure function is related to the valon
structure function by the convolution theorem as follows [1-5],
\b\ F_{2}^{N}(x,Q^{2})=\sum_{v}\int^{1}_{x}G_{\frac{v}{N}}(y)
F_{2}^{v}(\frac{x}{y},Q^{2})\frac{dy}{y}\e where the summation is
over the three valons, and $G_{\frac{v}{N}(y)}$ is the probability
of finding a valon that carrying momentum fraction y of the
hadron, and also  $F_{2}^{v}(\frac{x}{y},Q^{2})$ is the structure
function of a valon. The structure function of a U-type valon can
be written as:
 \b\
F^{U}_{2}(z,Q^{2})=\frac{4}{9}z(G_{\frac{u}{U}}+G_{\frac{\bar{u}}{U}})+\frac{1}{9}z(G_{\frac{d}{U}}+G_{\frac{\bar{d}}{U}}+G_{\frac{s}{U}}+G_{\frac{\bar{s}}{U}}),\e
where  G's  are valence  and sea quark distribution functions, and
defined   the probability function for $v$-valon  to have a
momentum fraction $z$ in the nucleon. A similar expression can be
written for the D-type valon. Structure of a valon can be written
in terms of favored distribution ($G_{f}$) and unfavored
distribution ($G_{nf}$), as for D and U valon structure functions
we have [1-5] \b\
F^{U}_{2}(z,Q^{2})=\frac{4}{9}z[G_{f}(z,Q^{2})+2G_{nf}(z,Q^{2})].\e
\b\ F^{D}_{2}(z,Q^{2})=
\frac{1}{9}z[G_{f}(z,Q^{2})+11G_{nf}(z,Q^{2})],\e or \b\
F^{U}_{2}(z,Q^{2})=\frac{2}{9}z[G^{S}(z,Q^{2})+G^{NS}(z,Q^{2})],\e
\b\
F^{D}_{2}(z,Q^{2})=\frac{1}{9}z[2G^{S}(z,Q^{2})-G^{NS}(z,Q^{2})],\e
where $G^{S}$and $G^{NS}$ are defined by singlet and nonsinglet
components respectively. The relation between favored and
unfavored distributions with singlet and nonsinglet distributions
have the following forms:\b\ G_{f}=\frac{1}{2f}[G^{S}
+(2f-1)G^{NS}],\e \b\ G_{uf}=\frac{1}{2f}[G^{S} -G^{NS}],\e where
f is number of active flavors (f=3 or f=4). In the momentum
representation we have \b\ M_{2}(n,Q^{2})=\int^{1}_{0} x^{n-2}
F_{2}(x,Q^{2}) dx,\e and \b\ M_{\alpha}(n,Q^{2})=\int^{1}_{0}
x^{n-1} G_{\alpha}(x,Q^{2}) dx \e where $\alpha=\frac{v}{N},S,NS$.
In order to estimate of the structure function momentums,
inserting Eqs.(9) and (10) in Eq. (1), as we obtain: \b\
M^{p}(n,Q^{2})=\sum_{v}M_{\frac{v}{p}}(n)M^{v}(n,Q^{2}).\e We
assume a general parameterization  form for U and V valons as
follows: \b\
G_{\frac{UUD}{p}}(y_{1},y_{2},y_{3})=\alpha(y_{1}y_{2})^{a}y_{3}^{b}\delta(y_{1}+y_{2}+y_{3}-1),\e
where a and b are the two free parameters that can be evaluated
from the experimental data, and also $\alpha$ is a normalization
coefficient. Here $y_{i}$ is the momentum fraction of the
i$^{,}$th valon. The $U$- and $D$-type valon distributions can be
obtained by integration over the specified variable as: \b\
G_{\frac{U}{p}}(y)=\int dy_{2}dy_{3}
G_{\frac{UUD}{p}}(y,y_{2},y_{3})=B(a+1,a+2,2)^{-1}y^{a}(1-y)^{a+b+1},\e
 \b\ G_{\frac{D}{p}}(y)=\int dy_{1}dy_{2}
G_{\frac{UUD}{p}}(y_{1},y_{2},y)=B(b+1,2a+2)^{-1}y^{b}(1-y)^{2a+1},\e
where $B(m,n)$ is the Euler beta- function. The normalization
factor has been fixed by requiring \b
\int_{0}^{1}G_{\frac{U}{P}}(y)dy=\int_{0}^{1}G_{\frac{D}{P}}(y)dy=1.
\e  Consequently, the moments of theses valon distributions are
calculated [1-6] according to the Mellin transformation from
Eq.(10) for nucleon \b\ U(n)=\frac{B(a+n,a+b+2)}{B(a+1,a+b+2)},\e
\b\ D(n)=\frac{B(b+n,2a+2)}{B(b+1,2a+2)},\e with $ a=0.65$ and
$b=0.35 $. Therefore, the valon  distributions can be obtained
as,\b\ G_{\frac{U}{p}}=7.98y^{0.65}(1-y)^{2},\e \b\
G_{\frac{D}{p}}=6.01y^{0.35}(1-y)^{2.3}.\e

  Now, we can go to the N-moment space for define the moments of these quark distribution functions (valence, sea , and
gluons)[5,7-8] as: \b\ M_{u_{v}}(n,s)=2U(n)M^{NS}(n,s),\e \b\
M_{d_{v}}(n,s)=D(n)M^{NS}(n,s),\e \b\
M_{sea}(n,s)=(2f)^{-1}[2U(n)+D(n)][M^{S}(n,s)-M^{NS}(n,s)],\e \b\
M_{g}(n,s)=[2U(n)+D(n)]M_{gq}(n,s),\e where $M^{S}$ and $M^{NS}$
are the moments of the singlet and nonsinglet valon structure
functions and $M_{gq}(n,s)$ is the quark-to-gluon evolution
function and defined into $d_{NS}$, $d_{+(-)}$, $d_{gq}$ and
$\rho$ where they  are anamolus dimensions [1-2,5,7-8] as follows:
 \b\ M^{NS}(n,s)=e^{-d_{NS}s},\e  \b\
 M^{S}(n,s)=\frac{1}{2}(1+\rho)e^{-d_{+}s}+\frac{1}{2}(1-\rho)e^{-d_{-}s},~~~~~~
  M_{gq}(n,s)=(d_{+}-d_{-})^{-1}d_{gq}[e^{-d_{+}s}-e^{-d_{-}s}].\e
Here, $s$ is defined by: \b\
s=\ln[\frac{\ln(\frac{Q^{2}}{\Lambda^{2}})}{\ln(\frac{Q_{0}^{2}}{\Lambda^{2}})}],\e
 where $Q^{2}_{0}$ and $\Lambda$ are our initial scales. In
 determining of the parton distributions, we have used a fit to a
 set of the experimental data [9-10] for a single value of $s$ or
 $Q^{2}$. Then we fit the moments by a beta function that are the
 moments of the forms [3-4,6]:
 \b\
xq_{v}(x)=a(1-x)^{b}x^{c},\e \b\ x{q}_{i}(x)=
a_{i}x^{b_{i}}(1-x)^{c_{i}}(1+d_{i}x+e_{i}x^{0.5}),\e where, the
subscript $i$ stand for $sea$ or $gluon$, and $xq_{i}$$^{,}$s are
the sea and gluon distribution functions. The free parameters in
Eqs.(27) and (28) are further considered to be functions of $s$,
as they are given in the  Appendix. Therefore we
 obtained the parton distributions for any valon that can be used
in the valon structure function.\\

\section*{3. Laguerre Polynomials to valon model}
So far, the structure of a nucleon into the valon distributions is
determined. Now we will use  an elegant and fast numerical method
for  determination of the proton structure function in valon
model. Therefore, we concentrate on the Laguerre polynomials in
our determinations. In the laguerre polynomials method [11-12],
the Laguerre polynomials are defined as: \b\
(n+1)L_{n+1}(x)=(2n+1-x)L_{n}(x)-nL_{n-1}(x),\e and orthogonality
condition is defined as: \b\
\int^{\infty}_{0}e^{-{\acute{x}}}L_{n}(\acute{x})L_{m}(\acute{x})d\acute{x}=\delta_{n,m}.\e
The general integrable function $f(e^{-\acute{x}})$ is transformed
into the sum: \b\
f(e^{-\acute{x}})=\sum^{N}_{0}f(n)L_{n}(\acute{x}),\e where \b\
f(n)=\int^{\infty}_{0}e^{-\acute{x}}L_{n}(\acute{x})f(e^{-\acute{x}})d\acute{x}.\e
In what follows we want to calculate the proton structure function
in valon model using the Laguerre polynomials method. We used the
variable transformations, $x=e^{-\acute{x}}$ , $y=e^{-\acute{y}}$
to get the valonic structure function form to the Laguerre
polynomials form. Then, we combined and expand each term of this
equation on Laguerre polynomials according to Eqs.(31)-(32) and
using this properties as: \b\int^{\acute{x}}_{0}d\acute{y}
L_{n}(\acute{x}-\acute{y})
L_{m}(\acute{y})=L_{n+m}(\acute{x})-L_{n+m+1}(\acute{x}).\e
 We obtained
an equation which determines $F^{p}_{2}(x,Q^{2})$ in terms of the
Laguerre polynomials, namely: \b F^{p}_{2}(n,Q^{2})=
\sum_{v}\sum^{n}_{m=0}\tilde{G}_{\frac{v}{p}}(m)[F^{v}_{2}(n-m,Q^{2})-F^{v}_{2}(n-m-1,Q^{2})],\e
where \b\ F^{v}_{2}(n,Q^{2})=\int^{\infty}_{0}d\acute{x}
e^{-\acute{x}}F^{v}_{2}(e^{-\acute{x}},Q^{2})L_{n}(\acute{x}),\e
and \b\tilde{G}_{\frac{v}{p}}(m)=\int^{\infty}_{0}d\acute{y}
e^{-\acute{y}}{G}_{\frac{v}{p}}(e^{-\acute{y}})L_{m}(\acute{y}),\e
as $F^{v}_{2}(x)$ is defined according to Eqs.(3) and (4) which
are accompanied with respect to Eqs.(27) and (28) and their
coefficients according to appendix, and also
${G}_{\frac{v}{p}}(y)$ is defined according to Eqs.(18) and (19)
respectively. Therefore we find the solution of the proton
structure function in valon model defined by solving this
recursion relation as: \b\
F^{p}_{2}(x,Q^{2})=\sum_{n=0}^{N}F^{p}_{2}(n,Q^{2})L_{n}(Ln\frac{1}{x}),\e
where $F^{p}_{2}(n,Q^{2})$  is  the proton structure function with
respect to the Laguerre model and its defined by Eqs.(34)-(36) as
the coefficients in these equations are obtained with respect to
the valon model. This result is completely general and gives the
expression for the proton structure function with respect to the
Laguerre polynomials model. Here we can expand the integrable
functions till a finite order $N=30$, as we can convergence these
series in the numerical
determinations.\\

\section*{4. Results and Discussion}
 We computed the predictions for all detail of the proton
structure function in the kinematic range where it has been
measured by $H1$ Collaboration [9-10] and compared with DL model
[13-15] based on hard Pomeron exchange, and with GJR
parametrization[16]. Our numerical predictions are presented as
functions of $x$ for the $Q^{2}=$22.5 $GeV^{2}$. The results are
presented in Fig.1 where they are compared with the
 $H1$ data and with the results obtained with the
help of other standard parameterizations.
The curves represent the
proton structure functions
  based on a fit to all data. We compared our results with
 predictions of $F_{2}^{p}$ in perturbative QCD  where
 the input densities are given by GJR parameterizations [16]. Also, we compared our results
   with the two pomeron fits as
seen in Fig.1. The agreement between the Laguerre polynomials
method for the proton structure function in valon model and data
at low and high -$x$ is remarkably good, as at low $x$ the gluon
distributions are dominate. Therefore, the good agreement
indicates that the Laguerre polynomials method in valon model for
the proton structure function has a good asymptotic behavior and
it is compatible  with both the data and the other standard models
at $x$ values. As this model has this advantage that we get a very
elegant solution for the proton structure function. Fig.2 shows
the shape of the distribution functions in Eqs.(27) and (28) for
the valence
and the sea quarks at $Q^{2}=22.5 GeV^{2}$.\\
In summary, we have used the Laguerre polynomials method to
describe the proton structure function in valon model. The proton
structure can be determined in terms of the valon distributions
and the valon structure functions with respect to Laguerre
polynomials. To confirm the method and results, the calculated
values are compared with the H1 data on the proton structure
function. It is shown that, there is a good agreement with
experimental H1 data for $F_{2}^{p}$, if one takes into the total
errors, and is consistent with a higher order QCD calculations of
$F_{2}^{p}$ which essentially show increase as $x$ decreases. We
observed that the calculations results are consistent with the two
pomeron model. Thus implying that Regge theory and perturbative
evolution may be made compatible at small-x. Also this model gives
a good description of the parton distributions at low
and high- $x$ values.\\
\section*{Appendix }
Here we will give the functional form of parameters of
Eqs.(27)-(28) by the following forms in terms of $s$.
 Coefficients for u valance in U valon are:
$$a_u=16.860-8.382s+1.352s^{2},$$
$$b_u=1.775+0.693s+0.014s^{2},$$
and $$c_u=1.794-0.699s+0.104s^{2}.$$ Coefficients for d valance in
D valon are:
$$a_d=13.119-8.66s+1.65s^{2},$$
$$b_d=7.129-3.5s+0.901s^{2},$$
 and $$c_d=1.971-0.9502s+0.163s^{2}.$$
 Coefficients for sea quarks
in each  valon are:
$$a_{sea}=-0.206+0.190s-0.022s^{2},$$
$$b_{sea}=-0.884+0.484s-0.104s^{2},$$
$$c_{sea}=12.089-4.532s+0.939s^{2},$$
$$d_{sea}=-2.564+5.937s-1.531s^{2},$$
and
 $$e_{sea}=-8.623+4.9823s-1.123s^{2}.$$
 Coefficients for gluons
in each  valon are:
$$a_{gluon}=13.8745-22.3304s+12.7885s^{2}-2.4801s^{3},$$
$$b_{gluon}=4.6810-8.4594s+4.7656s^{2}-0.9209s^{3},$$
$$c_{gluon}=-24.5652+50.4661s-30.147s^{2}+6.0738s^{3},$$
$$d_{gluon}=-0.8839+0.0403s-0.0174s^{2}.$$ \\
$\bf{Acknowledgements}$   G.R.Boroun would like to thank
Dr.M.Tabrizi for computer proceeding and Dr.H.Khanpour for fruitful discussions on QCD fits, and also Dr.B.Rezaei for reading and correcting the
manuscript of this paper and for productive discussions.\\

\begin{figure}
\includegraphics[width=1\textwidth]{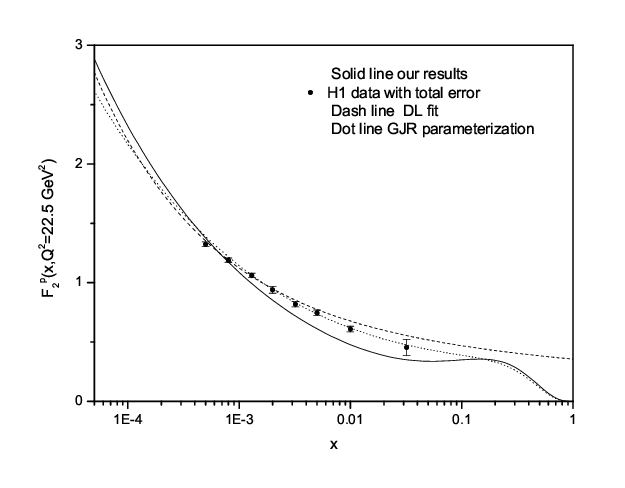}
\small
\begin{flushleft}
\caption{ Comparing of the proton structure function  by using the
Laguerre polynomials in valon model, with the experimental data
[9-10] and DL model [13-15] and GJR parameterization [16] at
$Q^{2}=22.5 GeV^{2}$.}
\end{flushleft}
\end{figure}
\begin{figure}
\includegraphics[width=1\textwidth]{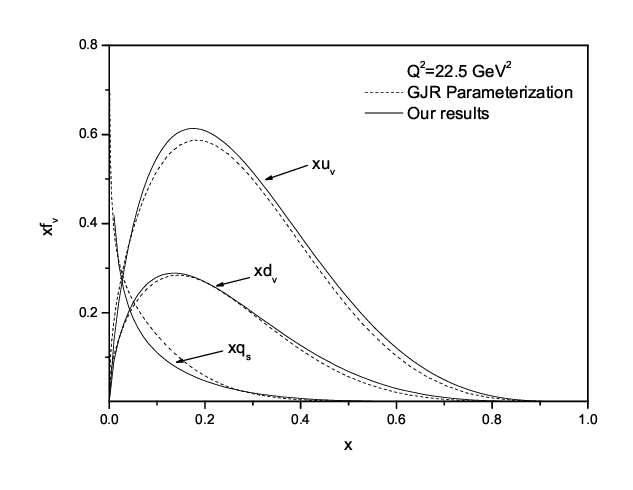}
\small
\begin{flushleft}
\caption{ Comparing of the parton distributions in proton  with
GJR parameterization [16] at $Q^{2}=22.5 GeV^{2}$.}
\end{flushleft}
\end{figure}

\begin{thebibliography}{a}
\bibitem{h01} R.C.Hwa, phys.Rev.D{\bf22}, 1593({1980}); phys.Rev.D{\bf51},
85({1995}).
\bibitem{h02} Rudolph Hwa and C.B.Yang, phys. Rev. C{\bf66},
025204({2002}); phys. Rev. C{\bf66}, 025205({2002}).
\bibitem{h03} Firooz Arash, arXiv:hep-ph/0307247V1,19Jul,{ 2003};Firooz Arash and Ali.N.khorramian,
 phys. Rev. C{\bf67}, 045201({2003}); Firooz Arash and
 Ali.N.khorramian, arXiv:hep-ph/990424V1,7Apr,{1999}; arXiv:hep-ph/9909328V1,11Sep,{
 1999}.
\bibitem{h04} Firooz Arash, Phys. Lett. B{\bf557}, 38({2003}); Phys.Rev. D{\bf 9}, 054024({2004}).
\bibitem{h05} R.C.Hwa and S.Zahir, Phy. Rev D,V{\bf23}, 2539({1981}); R.C.Hwa and C.S.Lam, Phy. Rev D,V{\bf26}, 2338({1982}).
\bibitem{h06} A.Mirjalili, et.al., J.Phys.G:Nucl.Part.Phys{\bf37}, 105003({2010}).
\bibitem{h07} T.A.Degrand,Nucl.Phys. B {\bf151},485({1979}).
\bibitem{h08} I.Hinchliffe and C.H.Llewellyn
Smith,B {\bf128}, 93({1977}).
\bibitem{h09} E665 Collab (Adams et al) Phys. Rev. Lett {\bf75},1466({1995});
H1 collab (Ahmed et al) Nucl. Phys. B{\bf439}, 471({1995}).
 \bibitem{h10} C.Adloff, H1 Collab., Eur.Phys.J.C{\bf21}, 33({2001}).
 \bibitem{h11} Laurent Schoeffel, Nucl.Instrum.Meth.A{\bf423}, 439({
 1999}); C.Coriano and C.Savkli, Comput.Phys.Commun.{\bf118},
 236(1999).
 \bibitem{h12} B.Rezaei and G.R.Boroun, Nucl.Phys.A {\bf857},  42({2011}).
\bibitem{h13} A.Donnachie and P.V.Landshoff, Phys. Lett.B {\bf296},  257({1992}).
\bibitem{h14} A.Donnachie and P.V.Landshoff, Phys. Lett.B {\bf437}, 408({1998}).
\bibitem{h15} A.Donnachie and P.V.Landshoff, Phys. Lett.B {\bf550},
160 ({2002});P.V.Landshoff ,arXiv:hep-ph/0203084.
\bibitem{h16}  M. Gluck, P. Jimenez-Delgado, E. Reya, Eur.Phys.J.C {\bf53},355
({2008}).

\end{thebibliography}
\end{document}